\documentclass{icrc2009}

\usepackage{graphicx}   
\usepackage[caption=false]{caption}    
\usepackage[font=footnotesize]{subfig} 
\usepackage{fixltx2e}
\usepackage{url}

\newcommand{\shorttitle}[1]%
{\markboth{Proceedings of the 31\MakeLowercase{$^{st}$} ICRC, {\L}\'{o}d\'{z} 2009}{#1} }
\newcommand{\etal}{\MakeLowercase{\textit{et al. }}} 

\begin{document}
\title{MWL observations of VHE blazars in 2006}

\author{\IEEEauthorblockN{S.\ R\"ugamer\IEEEauthorrefmark{1},
I.\ Oya\IEEEauthorrefmark{2},
M.\ Hayashida\IEEEauthorrefmark{3},
D.\ Mazin\IEEEauthorrefmark{4},
R.\ M.\ Wagner\IEEEauthorrefmark{5},
J.\ L.\ Contreras\IEEEauthorrefmark{2} and
T.\ Bretz\IEEEauthorrefmark{1}
\\
on behalf of the MAGIC collaboration}
\\
\IEEEauthorblockA{\IEEEauthorrefmark{1}Universit\"at W\"urzburg, D-97074 W\"urzburg, Germany}
\IEEEauthorblockA{\IEEEauthorrefmark{2}Universidad Complutense, E-28040 Madrid, Spain}
\IEEEauthorblockA{\IEEEauthorrefmark{3}SLAC National Accelerator Laboratory and KIPAC, CA, 94025, USA}
\IEEEauthorblockA{\IEEEauthorrefmark{4}IFAE, Edifici Cn., Campus UAB, E-08193 Bellaterra, Spain}
\IEEEauthorblockA{\IEEEauthorrefmark{5}Max-Planck-Institut f\"ur Physik, D-80805 M\"unchen, Germany}}

\shorttitle{S. R\"ugamer \etal blazar MWL observations}
\maketitle

\begin{abstract}
In 2006 the MAGIC telescope observed the well known very high energy (VHE, $>$\,80\,GeV) blazars Mrk\,421 and Mrk\,501 in the course of multi-wavelength campaigns, comprising measurements in the optical, X-ray and VHE regime.

MAGIC performed additional snapshot observations on Mrk\,421 around the MWL campaigns and detected the source each night with high significance, establishing once more flux variability on nightly scales for this object. For certain nights, the integral flux exceeded the one of Crab significantly, whereas the truly simultaneous observations have been conducted in a rather low flux state. The MAGIC observations contemporaneous to \textit{XMM-Newton} revealed clear intra-night variability. No significant correlation between the spectral index and the flux could be found for the nine days of observations.

The VHE observations of Mrk\,501 have been conducted during one of the lowest flux states ever measured by MAGIC for this object. The VHE and optical light curves do not show significant variability, whereas the flux in X-rays increased by about 50\,\%.

In this contribution, the results of the MAGIC observations will be presented in detail.
\end{abstract}


\section{Introduction}
Blazars are radio-loud active galactic nuclei (AGN) viewed at small angles between the jet axis and the line of sight. Due to relativistic beaming, they are the brightest and most variable high energy sources among AGNs. BL Lacertae objects, a subclass of blazars \cite{Urry96}, are completely dominated by nonthermal emission and show only faint or no emission lines at all. Their spectral energy distribution (SED) extends over more than 15 orders of magnitude and exhibits two pronounced peaks when plotted as $\nu F_{\nu}$ versus $\nu$. High-peaked BL Lacertae objects (HBLs) are characterized by the first peak being located in the UV to X-ray range, whereas the second peak appears at $\gamma$-ray energies (GeV to TeV).

The non-thermal, beamed emission detected from HBLs is commonly ascribed to synchrotron and inverse-Compton radiation from ultrarelativistic charged particles, accelerated at shocks in jets moving with relativistic bulk speed. Progress in understanding the roles of shocks and plasma turbulence for particle acceleration and the interplay with the dynamics of the jet expansion is expected from modelling the SED of blazars. At present, it is unclear whether the observed emission is entirely due to accelerated electrons, as in Synchrotron-Self-Compton (SSC) models (e.g.\ \cite{Marscher85}, \cite{Maraschi92}, \cite{Costamante02}), or whether the high-energy emission is due to pions produced by accelerated protons and ions and subsequent pion decay (e.g.\ \cite{Mannheim93}, \cite{Muecke03}).

Owing to their strong spectral and flux variability at all wavebands down to minutes (see e.g.\ \cite{Mrk501flare}), simultaneous multi-wavelength (MWL) campaigns are essential to study the SED of HBLs. Of paramount importance are hereby the X-ray and $\gamma$-ray regimes, where the two SED peaks are located for HBLs. In former MWL campaigns, the minimum time scale of truly simultaneous observations was restricted by the low sensitivity of the participating VHE telescopes, enabling the investigation of transient events only in high flux states. With the advent of a new generation of imaging atmospheric Cherenkov telescopes with improved sensitivity and lowered energy threshold such as MAGIC, H.E.S.S.\ and VERITAS, the detection of HBLs in a few hours becomes feasible also during low flux states.

In this proceedings we present MAGIC (\cite{Baixeras04}, \cite{Cortina05}) observations of the well-known TeV blazars Mrk\,421 and Mrk\,501 in 2006. For individual nights, these were conducted (quasi-)simultaneously with H.E.S.S.\ \cite{Aharonian06}, Suzaku \cite{Mitsuda07}, \textit{XMM-Newton} \cite{Jansen01}, INTEGRAL \cite{Winkler03} and KVA \cite{Takalo08}.

\begin{table*}
\caption{ Analysis results for Mrk\,421 (April and June 2006).}
\label{tab:results}
\centering
\begin{tabular}{ccccccccc}
\hline (1) & (2) & (3) & (4) & (5) & (6) & (7) & (8) & (9) \\
Observation Night & $S$ & Time [h] & $E_\mathrm{min}$ [GeV] & $F(E>E_\mathrm{min})$ & $\chi2_\mathrm{red,const}$ & $F_0$ & $\alpha$ & $\chi2_\mathrm{red,PL}$ \\
\hline \hline
2006/04/22 &           10.9\,$\sigma$ & 0.76 & 250 & $0.917\pm0.112$                                        & \phantom{0}1.28/2 & $2.37\pm0.33$ & $2.05\pm0.23$ & 2.07/2 \\
2006/04/24 &           25.0\,$\sigma$ & 0.99 & 250 & $\phantom{0}2.32\pm0.13\phantom{0}$ & \phantom{0}2.71/2 & $5.14\pm0.40$ & $2.25\pm0.09$ & 1.95/3 \\
2006/04/25 &           20.8\,$\sigma$ & 1.53 & 250 & $\phantom{0}1.34\pm0.09\phantom{0}$ & \phantom{0}1.67/2 & $2.99\pm0.31$ & $2.26\pm0.12$ & 0.24/3 \\
2006/04/26 &           16.4\,$\sigma$ & 1.64 & 250 & $\phantom{0}1.08\pm0.09\phantom{0}$ & \phantom{0}1.32/4 & $2.37\pm0.26$ & $2.35\pm0.17$ & 0.41/2 \\
2006/04/27 &           33.5\,$\sigma$ & 1.42 & 250 & $\phantom{0}3.21\pm0.15\phantom{0}$ & \phantom{0}1.87/4 & $8.05\pm0.42$ & $2.07\pm0.07$ & 4.79/4 \\
2006/04/28 &           19.9\,$\sigma$ & 2.23 & 250 & $\phantom{0}1.14\pm0.08\phantom{0}$ & \phantom{0}4.34/8 & $2.38\pm0.22$ & $2.47\pm0.14$ & 0.65/2 \\
2006/04/29 &           23.7\,$\sigma$ & 2.78 & 250 & $\phantom{0}1.04\pm0.06\phantom{0}$ &                       41.0/7 & $2.35\pm0.17$ & $2.28\pm0.09$ & 2.04/4 \\
2006/04/30 &           10.3\,$\sigma$ & 0.16 & 250 & $\phantom{0}2.39\pm0.33\phantom{0}$ &                             --- & $6.85\pm1.06$ & $1.66\pm0.18$ & 1.36/1 \\
2006/06/14 & \phantom{0}7.5\,$\sigma$ & 0.80 & 450 & $0.341\pm0.059$                               & \phantom{0}2.36/1 & $1.68\pm0.32$ & $2.38\pm0.44$ & 1.45/2 \\
\hline
\end{tabular}

(1) Day at which the observations started, (2) resulting significances $S$, (3) effective observation time, (4) minimum spectral energy $E_\mathrm{min}$ used for the calculation of the integral fluxes, (5) integral fluxes $F$ above $E_\mathrm{min}$ (in units of $10^{-6}\mathrm{m}^{-2} \mathrm{s}^{-1}$), (6) fit quality of a constant-flux fit to the individual observation nights, (7)\,-\,(9) power-law fit results for the differential energy spectra of ${\mathrm d}N/{\mathrm d}E=F_0 \cdot (E/E_0)^{-\alpha}$ with $E_0=1\,\mathrm{TeV}$; $F_0$ in units of $10^{-7}\mathrm{m}^{-2} \mathrm{s}^{-1}$.
\end{table*}

Mrk\,421 (z\,=\,0.030) was the first extragalactic object detected at VHE energies \cite{Punch92}. Since then, various studies from radio wavelength up to VHE $\gamma$-rays have shown strong variability patterns for this object. Flux variations by more than one order of magnitude (e.g.\ \cite{Fossati08}), and occasional flux doubling times as short as 15 minutes (\cite{Gaidos96}, \cite{Aharonian02}, \cite{Schweizer08}) have been reported. Variations in the hardness of the TeV $\gamma$-ray spectrum during flares were reported by several groups (e.g.\ \cite{Krennrich02}, \cite{Aharonian05}, \cite{Fossati08}). Simultaneous observations in the X-ray and VHE bands show strong evidence for correlated flux variability (\cite{Krawczynski01}, \cite{Blazejowski05}, \cite{Fossati08}).

The second established TeV blazar \cite{Quinn96} Mrk\,501 (z\,=\,0.034) has been classified as an 'extreme' HBL, having shown a shift of the synchrotron peak of about two orders of magnitude to energies higher than 100\,keV during observation of a flare by BeppoSAX in 1997 (\cite{Catanese97}, \cite{Pian98}, \cite{Tavecchio01}). Also at VHE energies surprisingly high activity (up to 10 times the Crab Nebula flux) has been detected at that time \cite{Aharonian01}. Reference \cite{Gliozzi06} describes a long term monitoring campaign in 2004, also covering the X-ray and TeV energy bands. It confirms the presence of a direct correlation between X-ray and VHE $\gamma$-ray emission, which appears to be stronger when the source is brighter. In 2005, when MAGIC observed the object between May and July, the source flux varied by an order of magnitude. During the two most active nights, rapid flux variability with a doubling time of a few minutes was observed \cite{Mrk501flare}. Several extensive SED studies based on MWL observations of this object were reported (e.g.\ \cite{Kataoka99}, \cite{Krawczynski00}, \cite{Sambruna00}, \cite{Tavecchio01}, \cite{Ghisellini02}). However, these observations only covered high flux states, rendering the statistics of low state MWL studies sparse.

\section{Observations and Data Analysis}
All MAGIC data were taken during dark nights in wobble mode \cite{wobble}. Observations of Mrk\,421 have been conducted by MAGIC in the nights from MJD 53848 through 53856 (2006 April 22$^{\mathrm{nd}}$\,-\,30$^{\mathrm{th}}$), accompanied by simultaneous KVA observations for most of the nights. From MJD 53854.87 - 53855.35 and from 53853.28 - 53854.27 Suzaku and \textit{XMM-Newton}, respectively, provided additional simultanous X-ray coverage. In the Suzaku window, H.E.S.S.\ conducted observations previous to the MAGIC ones, as did Whipple 3.5\,h after the MAGIC observations in the \textit{XMM-Newton} time slot. Apart from these dedicated campaigns, MAGIC and KVA took data on the source at MJD 53901, following a target-of-opportunity alert from INTEGRAL. Altogether, the MAGIC pointings resulted in 11.9\,h effective observation time in April with a zenith distance (zd) range of 9$^\circ$\,-\,41$^\circ$ and 0.80\,h for 43$^\circ$\,$<$\,zd\,$<$\,52$^\circ$ in June. The detection significances, calculated using formula 17 of \cite{LiMa}, were 64.8\,$\sigma$ and 7.5\,$\sigma$ for the April and June dataset, respectively (see Table \ref{tab:results}). For more details see \cite{Mrk421this}.

The MWL campaign on Mrk\,501 started on the 18$^{\mathrm{th}}$ of July 2006 (MJD 53935). The nightly MAGIC, H.E.S.S.\ and KVA observations (MJD 53935 - 53937) where covered by a continuous Suzaku pointing from MJD 53934.77 - 53935.73. The 9.1\,h of MAGIC effective on-time (11$^\circ$\,$<$\,zd\,$<$\,35$^\circ$) for all three nights resulted in a detection of Mrk\,501 with 13.4\,$\sigma$ significance. Observation details can be found in \cite{Mrk501this}.

The data were processed using the analysis and reconstruction software package for MAGIC data (\cite{CrabMAGIC}, \cite{MarsW}). Detailed information on the different analysis steps and performance are given in \cite{CrabMAGIC}. The MAGIC observations on Mrk\,421 presented here are among the first data taken after major hardware updates in April 2006, causing a higher than usual energy threshold (see \cite{Mrk421this}).

\section{Results}

\begin{figure}[!t]
\centering
\includegraphics[width=3in]{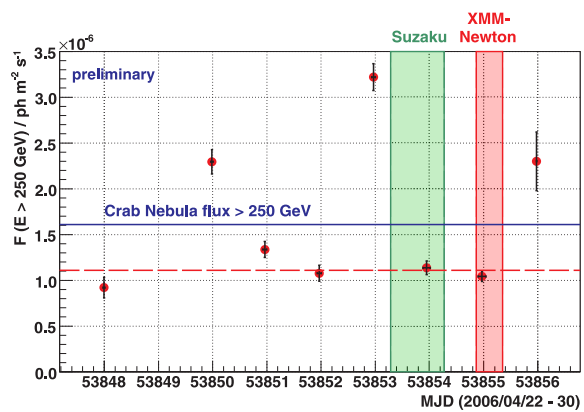}
\caption{MAGIC LC (E\,$>$\,250\,GeV) for the Mrk\,421 observations in April. The Crab Nebula flux as measured by MAGIC is shown as a blue solid line. The dashed red line corresponds to the average flux during the low states. Also shown are the observations windows of Suzaku (green bar) and \textit{XMM-Newton} (red bar).}
\label{fig:LC_inter}
\end{figure}

MAGIC detected the sources each night with high significance. Independent analyses confirm the results presented here.

\subsection{Mrk\,421}
The results of the daily data analysis are summarized in Table \ref{tab:results}. For Mrk\,421, not all resulting spectra for the individual nights were consistent with a simple power law. For MJD 53853 a likelihood ratio test prefered a log-parabolic power law \cite{Massaro04} or a power law with an exponential cut-off over a simple power law with a probability of 94.9\,\% and 95.2\,\%, respectively. Also combining all data from April, the likelihood ratio probability clearly showed evidence of a parabolic or cut-off shape of the spectrum.

\begin{figure}[!t]
\centering
\includegraphics[width=3in]{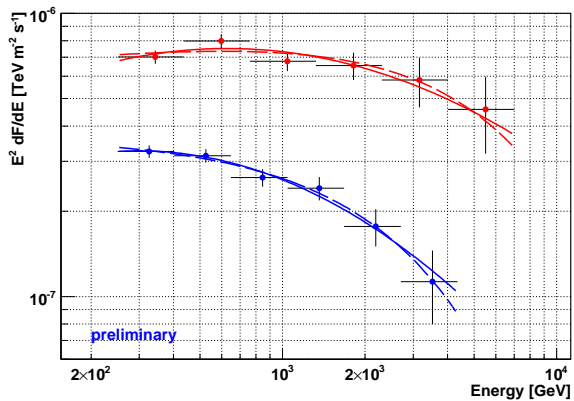}
\caption{Differential energy spectra times E$^2$ for the combined low (blue) and high (red) state nights of Mrk\,421. The data were fitted using a log-parabolic power law (solid line) and a power law with an exponential cut-off (dashed line).}
\label{fig:spectra_hilo}
\end{figure}

The nocturnal light curve (LC) for the April measurements is shown in Figure \ref{fig:LC_inter}. For three nights the integral flux of Mrk\,421 was higher than the one of the Crab Nebula, significantly exceeding the flux level of the five other nights. The average flux in the high state nights (deduced from a signal of 46.4\,$\sigma$) was (2.75\,$\pm$\,0.09)\,$\times$\,10$^{-6}$\,m$^{-2}$\,s$^{-1}$, more than twice the average flux of the low state nights ((1.09\,$\pm$\,0.03)\,$\,\times$\,10$^{-6}$\,m$^{-2}$\,s$^{-1}$ from 44.6\,$\sigma$). The combined high and low flux spectra are shown in Figure \ref{fig:spectra_hilo}. Both could not be described sufficiently by a simple power law, but were modelled equally well by a log-parabolic power law and a power law with exponential cut-off, respectively, yielding frequencies for the position of the second SED peak of (297\,$\pm$\,176)\,GeV and (627\,$\pm$\,187)\,GeV (log-parabolic) and $<$\,250\,GeV and (603\,$\pm$\,850)\,GeV (cut-off) for the low and high state, respectively. Additionally, the latter model allowed us to derive the cut-off energy of the spectrum, located at (3.1\,$\pm$\,1.5)\,TeV for the low state nights and at (6.3\,$\pm$\,4.2)\,TeV for the high state nights.

\begin{figure}[!t]
\centering
\includegraphics[width=3in]{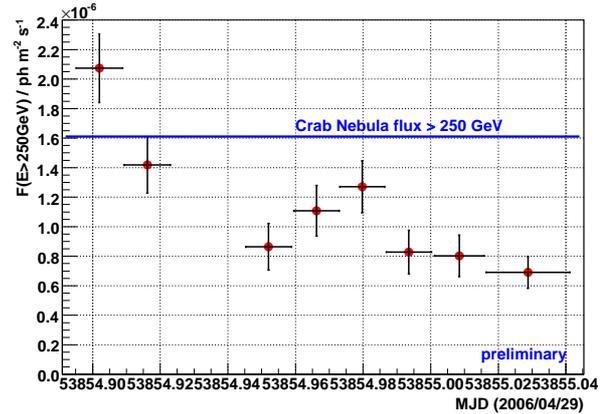}
\caption{Mrk\,421 intra-night MAGIC LC (E\,$>$\,250\,GeV) for the night of the 29$^{\mathrm{th}}$ of April. The Crab Nebula flux as measured by MAGIC is shown as a blue line.}
\label{fig:LC_intra}
\end{figure}

Thanks to the high significance detections each night we were able to derive the VHE LCs for the individual nights on sub-hour scale. MAGIC found a clear indication of intra-night variability for the observations simultaneous to the \textit{XMM-Newton} pointings, as shown in Figure \ref{fig:LC_intra}. A fit to a constant yielded an unacceptable probability of 8.0\,$\times$\,10$^{-5}$\,\%. The LCs of the other nights were consistent with a constant flux (see \cite{Mrk421this}).

\begin{figure}[!t]
\centering
\includegraphics[width=3in]{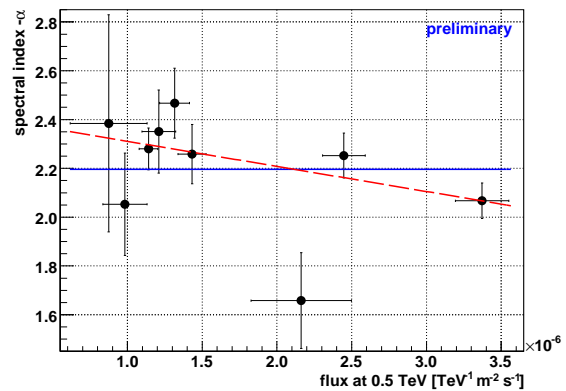}
\caption{Measured flux at 500\,GeV versus spectral index for all Mrk\,421 measurements of this campaign. The blue solid line shows a fit of a constant to the data, the dashed red one a linear fit.}
\label{fig:correlation}
\end{figure}

In Figure \ref{fig:correlation} the flux at 500\,GeV is plotted versus the spectral index for all nine nights of MAGIC observations of Mrk\,421. A significant correlation of the flux with the spectral index is not apparent but cannot be ruled out either ($\chi2/dof_{const} = 17.3/8$, $\chi2/dof_{linear} = 11.5/7$).

\subsection{Mrk\,501}
The average spectrum of Mrk\,501 measured by MAGIC in this campaign could be well described by a simple power law from 80\,GeV to 2\,TeV with a photon index of $2.79 \pm 0.12$. The integral flux above 200\,GeV was (4.6\,$\pm$\,0.4)\,$\times$\,10$^{-7}$\,m$^{-2}$\,s$^{-1}$, corresponding to $\sim$\,23\,\% of the Crab Nebula flux as measured by MAGIC.

\begin{figure}[!t]
\centering
\includegraphics[width=3in]{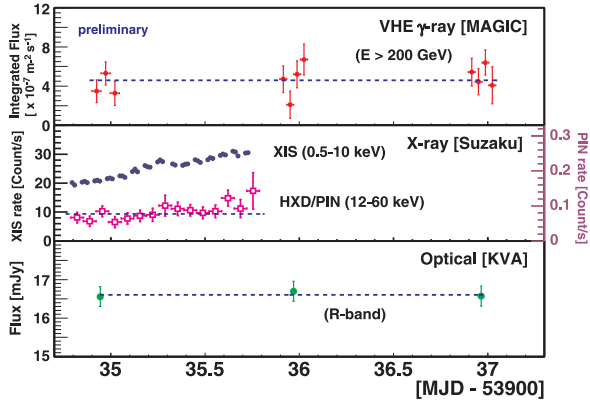}
\caption{MAGIC (top), Suzaku (middle) and KVA (bottom) LC for Mrk\,501 derived in this campaign. The dashed horizontal lines represent the average fluxes of the measurements.}
\label{fig:LC_Mrk501}
\end{figure}

The light curves obtained from the MWL campaign by MAGIC, Suzaku and KVA are shown in Figure \ref{fig:LC_Mrk501}. The measured flux in the VHE regime did not vary significantly and was comparable with the lowest value of MAGIC measurements in 2005 and 2006. The Suzaku XIS data showed a flux increase of about 50\,\% during the observations.

\section{Discussion}
The well-know TeV blazars Mrk\,421 and Mrk\,501 have been observed by MAGIC in the course of several MWL campaigns. Most of the simultaneous observations were prearranged and catched the sources in rather low flux states, adding statistics to the small sample of truly contemporaneous low state data.

Whereas the monitoring of Mrk\,421 througout the campaigns exhibited strong variability between the individual observation nights as well as intra-night, the observations on Mrk\,501 revealed no significant varability. The intra-night variability pattern found for Mrk\,421 did not show up in the contemporaneous X-ray data by \textit{XMM-Newton} (see \cite{Mrk421XMM}). On the other hand, the flux increase of Mrk\,501 in the soft X-rays could not be found in the VHE data. However, in this second case, due to the low source flux level, MAGIC could only have seen variability if the flux were to increase by a factor of 2-3, a level of change which was also not apparent in X-rays.

Using the results of the nine days of MAGIC observations on Mrk\,421 we searched for correlation between the fluxes and spectral indices. No significant correlation was apparent, which may be due to the small dynamical range of the parameters found here.

Dividing the observations in a low and high state data set, we found that both spectra were not compatible with a simple power law but were well fit by a log-parabolic power law and also a power law with exponential cut-off. Our data did not allow to prefer one model over the other. If source-intrinsic, a log-parabolic spectral shape in the VHE regime can be explained by stochastic, energy-dependent acceleration of electrons within the SSC framework \cite{Massaro04}. Electron scattering in the Klein-Nishina regime or absorption of the $\gamma$-rays by extragalactic background photon fields (e.g.\ \cite{Kneiske2008}) can be responsible for an intrinsic or cosmological spectral cut-off feature, respectively.

The peak frequencies dervied from both fits were compatible with each other and showed indications for an increase of the peak energy with rising flux level, as predicted if the VHE radiation were due to SSC mechanisms. The derived cut-off energies were well below the limit of $\sim$\,13\,TeV deduced from the Kneiske lower limit model \cite{Kneiske2008}, indicating a source-intrinsic rather than a cosmological reason for the cut-off feature.

A detailed discussion of the data in a MWL context will be presented in subsequent publications, but some aspects can already be found in \cite{Mrk421XMM} and \cite{Mrk501this}.

\vspace{1cm}
\textit{Acknowledgments.}
We would like to thank the Instituto de Astrofisica de Canarias for the excellent working conditions at the Observatorio del Roque de los Muchachos in La Palma. The support of the German BMBF and MPG, the Italian INFN and Spanish MICINN is gratefully acknowledged. This work was also supported by ETH Research Grant TH 34/043, by the Polish MNiSzW Grant N N203 390834, and by the YIP of the Helmholtz Gemeinschaft. S.\ R.\ is supported by Research Training Group 1147 ``{\it Theoretical Astrophysics and Particle Physics}'' of Deutsche Forschungsgemeinschaft.

\end{document}